\documentclass[sigconf]{acmart}
\usepackage{balance}
\usepackage{url}
\usepackage{booktabs} % For formal tables

\usepackage{soul,color}

\usepackage{graphicx}
\usepackage{subfig}
\usepackage{tabularx}

\usepackage{amsmath}
\usepackage{amssymb}
\usepackage{algorithm}
\usepackage[noend]{algpseudocode}
\usepackage{paralist}

\setcopyright{rightsretained}

\begin{document}

\copyrightyear{2020}
\acmYear{2020}
\acmConference[SPAA '20]{Proceedings of the 32nd ACM Symposium on Parallelism in Algorithms and Architectures}{July 15--17, 2020}{Virtual Event, USA}
\acmBooktitle{Proceedings of the 32nd ACM Symposium on Parallelism in Algorithms and Architectures (SPAA '20), July 15--17, 2020, Virtual Event, USA}
\acmDOI{10.1145/3350755.3400266}
\acmISBN{978-1-4503-6935-0/20/07}

\title{Brief Announcement: On the Limits of Parallelizing Convolutional Neural Networks on GPUs}

\author{Behnam Pourghassemi}
\email{bpourgha@uci.edu}
\affiliation{
  \institution{University of California, Irvine}
 }
 
 \author{Chenghao Zhang}
\email{chenghz4@uci.edu}
\affiliation{
	\institution{University of California, Irvine}
}

\author{Joo Hwan Lee}
\email{joohwan.lee@samsung.com}
\affiliation{
	\institution{Samsung Semiconductor}
}
  
\author{Aparna Chandramowliswharan}
\email{amowli@uci.edu}
\affiliation{
	\institution{University of California, Irvine}
}

\begin{abstract}
GPUs are currently the platform of choice for training neural networks. 
However, training a deep neural network (DNN) is a time-consuming process even on GPUs because of the massive number of parameters that have to be learned.
As a result, accelerating DNN training has been an area of significant research in the last couple of years. 

While earlier networks such as AlexNet had a linear dependency between layers and operations, state-of-the-art networks such as ResNet, PathNet, and GoogleNet have a non-linear structure that exhibits a higher level of inter-operation parallelism.
However, popular deep learning (DL) frameworks such as TensorFlow and PyTorch launch the majority of neural network operations, especially convolutions, serially on GPUs and do not exploit this inter-op parallelism.
In this brief announcement, we make a case for the need and potential benefit of exploiting this rich parallelism in state-of-the-art non-linear networks for reducing the training time.
We identify the challenges and limitations in enabling concurrent layer execution on GPU backends (such as cuDNN) of DL frameworks and propose potential solutions.
\end{abstract}

%
% The code below should be generated by the tool at
% http://dl.acm.org/ccs.cfm
% Please copy and paste the code instead of the example below.
%




\maketitle

\section {Introduction}
\begin{figure}[htbp]
\centering
  \includegraphics[width = 0.42\textwidth]{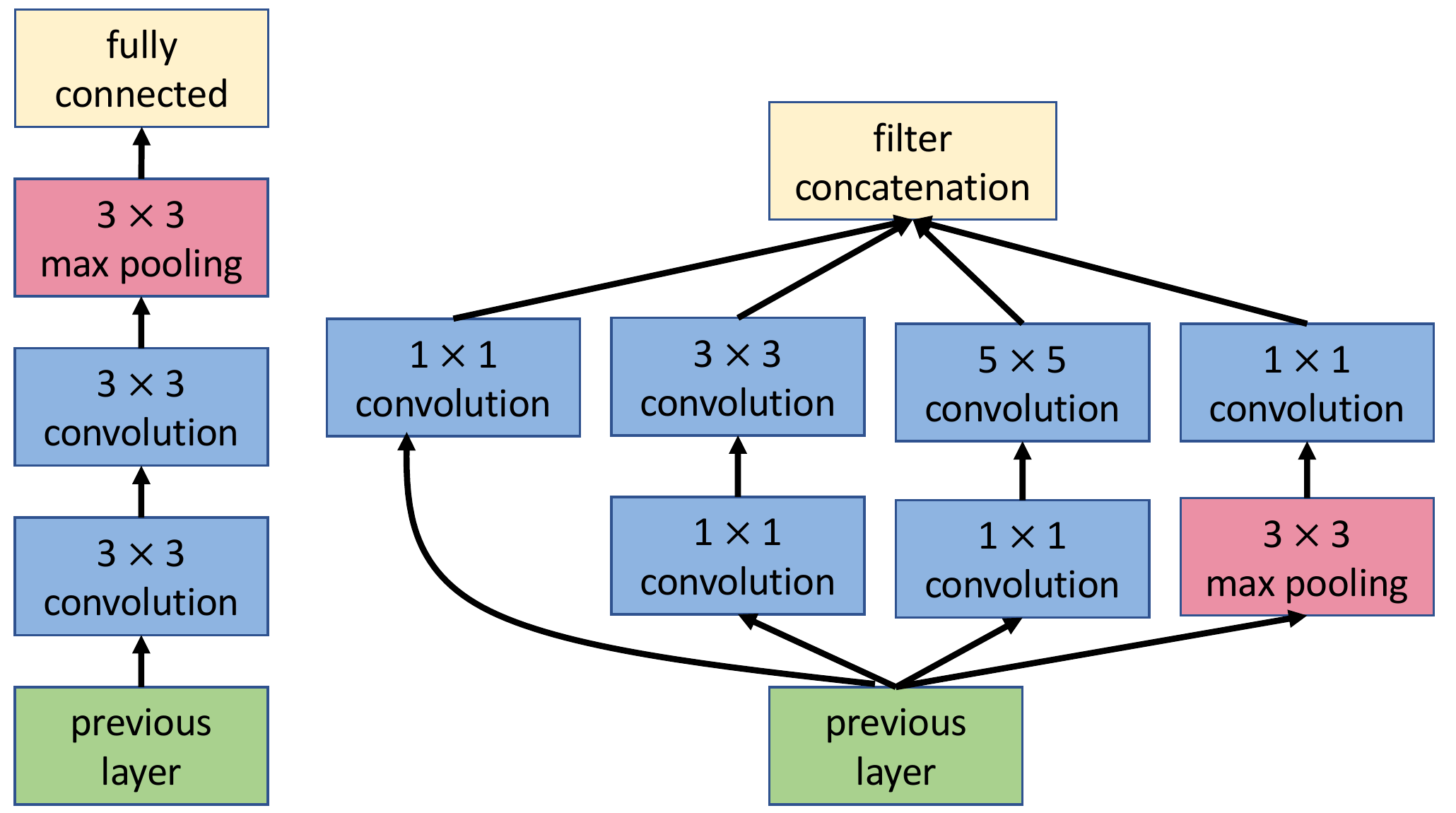}
  \caption{Examples of linear (AlexNet on the left) and non-linear (GoogleNet on the right) networks.} \label{fig:network}
   \end{figure}

Convolutional Neural Networks (CNNs) are a popular class of DNNs with several applications such as computer vision \cite{vision, vgg}, voice recognition \cite{speech}, recommender systems \cite{recommender}, physics simulations \cite{obiols2020cfdnet}, and natural language processing \cite{nlp,nlp2}.
Earlier CNNs were composed of a \emph{linear} sequence of dependent layers like VGG and AlexNet.
However, modern networks such as ResNet, GoogleNet, DenseNet, and PathNet have a more complex architecture.
These \emph{non-linear} networks contain multiple fork/joins resulting in independent paths of chained operations.
Figure~\ref{fig:network} illustrates the difference in structure between linear (AlexNet) and non-linear (GoogleNet) networks.

GPUs are the platform of choice for training CNNs.
Training large-scale CNNs is extremely time-consuming due to the ever-growing number of parameters that have to be learned and the numerous iterations for the model to converge.
Two approaches to reducing training time are to increase throughput and reduce the per-iteration execution time.
For the former, it is common to parallelize training on multiple GPUs or GPU clusters using different strategies \cite{jia2018beyond, ben2019demystifying}.
For the latter, however, there are several solutions in literature \cite{graphi}.
They can be broadly classified into either optimizing the operations in each layer or exploiting the concurrency between CPUs and GPUs by pipelining pre-processing operations (such as resizing, normalization) on the CPU with the rest of the operations on the GPU \cite{superneurons, cnnperformance}. 
As one can infer from Figure~\ref{fig:network}, unlike linear networks, non-linear networks have multiple independent operations across layers. 
However, none of the above state-of-the-art approaches exploit this parallelism across multiple paths by running independent operations across layers concurrently on a single GPU.
In this paper, we investigate \emph{why and how} to utilize this rich inter-op parallelism in non-linear CNNs to reduce training time. 

\section{Parallel convolutions on a GPU}

\begin{table*}[htbp]
%\begin{figure*}[htbp]
\begin{tabular}{|c|c|c|c|c|c|c|c|c|}

 \hline
 Layer               &  Algorithm  & Kernel name                       &Registers     & Shared Memory &  Threads      &  Blocks         &  ALUs     &  Memory stalls \\
   \hline
   \hline
 Incep. 1    &   PRECOMP\_GEMM          & implicit\_convolve\_sgemm  &    $92\%$   &    $39\%$    &    $38\%$    &    $19\%$     & $70\% $ &     $0.47\%$  \\
 %\cline{2-9}
 ($3*3$)    &   FFT\_TILING                    & fft2d\_c2r\_32x32                  &    $38\%$   &    $75\%$    &    $25\%$    &    $6\%$       & $30\% $ &     $15.2\%$  \\
     \hline
Incep. 1    &   PRECOMP\_GEMM        & implicit\_convolve\_sgemm  &    $100\%$  &    $70\%$   &    $50\%$    &   $100\%$    & $60\% $ &     $0.03\%$  \\
% \cline{2-9}
 ($5*5$)    &  FFT\_TILING                     & fft2d\_c2r\_32x32                 &    $38\%$   &    $75\%$    &    $25\%$    &    $6\%$       & $20\% $ &     $16.5\%$  \\
 
 \hline
\end{tabular}
\vspace{0.4cm}
 \caption{Resource utilization of two different algorithms for two independent convolutions in the first Inception module of GoogleNet on a Tesla K40 GPU.}\label{table:sm}
\end{table*} 
%\end{figure*}

A majority of DL frameworks have a GPU backend that compiles the model and generates a computation graph at the granularity of basic operations such as convolution, batch normalization, and pooling.
The operations are executed on the device by calling the corresponding APIs in \emph{highly optimized} third-party libraries such as Nvidia cuDNN \cite{cudnn} and cuBLAS \cite{cublas}.
The kernels implemented in these libraries hold device resources to perform the CNN operations.

The core operation in CNNs is \emph{convolution} which constitutes the majority of the training time, approximately 60\% of the compute time for ImageNet Large Scale Visual Recognition Challenge (ILSVRC) winners \cite{superneurons}.
It also typically consumes more memory than other network layers \cite{cnnperformance, superneurons}.
cuDNN supports multiple algorithms for each type of convolution. 
For example, for forward convolution, it supports \texttt{GEMM}, \texttt{IMPLICIT\_GEMM}, \texttt{IMPLICIT\_PRECOMP\_GEMM}, \texttt{WINOGRAD}, \texttt{WINOGRAD\_NONFUSED}, \texttt{DIRECT}, \texttt{FFT}, and \texttt{FFT\_TILING}. 
Depending on the convolution parameters (input, filter, data layout, etc.), each of the above algorithms has a different execution time, resource utilization, and workspace memory.
  
To launch multiple convolutions concurrently on a GPU, each convolution has to be assigned to a separate executor (\emph{stream} in the CUDA programming model).
Besides, to accommodate two or more convolutions on a GPU, DL frameworks need to ensure there is enough device memory available at launch time \footnote{Even though CUDA unified memory can use CPU memory, the communication cost can outweigh the parallelization payoff.}. 
Convolutions in cuDNN use device global memory for storing input, output, filter, and intermediate results (or workspace). 
The input, output, and filter sizes for convolutions are fixed during model construction, so DL frameworks can only adjust workspace memory. 

\subsection{Results and Analysis}
Our experiments on numerous convolutions (from popular networks such as GoogleNet and ResNet) reveal that it is not feasible to run two or more cuDNN convolutions concurrently.
Using the Nvidia profiler, we observe that cuDNN kernels exhaust one or more resources such as registers and shared memory on the GPU Streaming Multiprocessor (SM) and do not allow the GPU scheduler to execute blocks from another kernel on the same SM.
Since a convolution typically has enough blocks to occupy all available SMs, execution of a second convolution is postponed to after the first convolution is completed resulting in a sequential execution of the two operations. 
Even though the profiler reports high occupancy for convolutions, for combinations of inputs and convolution algorithms, the computational efficiency and DRAM utilization are not high enough (e.g. less than 50\%) \cite{maxdnn, analyzingML, li2016optimizing, lavin2016fast}.

In addition, current DL frameworks either stick to certain algorithms for convolutions or pick the fastest algorithm.
For example, in the first iteration, TensorFlow (r1.10) tests all algorithms for each convolution and chooses the fastest one for subsequent iterations. 
Even though this method is optimal to reduce the execution time of linear networks, it is not essentially the best option for the parallel execution of operations since the fastest algorithm could inadequately use SM resources and/or consume a large amount of workspace memory preventing concurrent kernel executions. 
We observe this exact behavior by profiling the resource utilization and workspace memory of convolutions in popular networks. 

\textbf{\emph{SM resources.}}
Table \ref{table:sm} shows profiling data for two independent convolutions in the inception module of GoogleNet on a Tesla K40 GPU with CUDA 10.0 and cuDNN 7.6.
According to the table, PRECOMP\_GEMM algorithm for the first convolution exhausts SM registers (more than 90\%) but poorly uses shared memory (39\%) while FFT\_TILING algorithms have complementary static resource utilization, i.e. bottlenecked by SM shared memory but consume only $38\%$ of registers. 
Further, these two algorithms exhibit different warp execution characteristics. 
For example, FFT\_TILING (on the second convolution) has 20\% ALU utilization but significantly greater memory stalls compared to the PRECOMP\_GEMM algorithm (on the first convolution) with high ALU utilization (70\%) and lower memory stalls. 
This indicates, the former algorithm is relatively bound by memory rather than compute resources as in the latter algorithm. 

In the past few years, researchers have proposed inter-SM \cite{adriaens2012case, zhao2018classification, park2017dynamic} and intra-SM \cite{dai2018accelerate,xu2016warped,wang2016simultaneous, park2017dynamic} partitioning to improve resource utilization for concurrent kernel execution. 
Inter-SM partitioning or spatial multitasking \cite{adriaens2012case, zhao2018classification} which partitions the SMs among kernels has performance benefits when kernels with complementary characteristics are co-located. In intra-SM partitioning, resource utilization is further improved by letting blocks from different kernels share the same SM. For instance, functional units in an SM (ALUs, SFUs, etc.) that are idle when running a memory-intensive kernel can be utilized by the blocks of a compute-intensive kernel. Intra-SM partitioning can practically be achieved when one or more SM static resources such as registers and shared memory remain under-utilized by kernels \cite{dai2018accelerate, xu2016warped, park2017dynamic}

  \begin{table}[htbp]
\begin{tabular}{|c|c|c|}

 \hline
 Convolution Algorithm & Workspace Memory & Runtime \\
  \hline
   \hline
   GEMM & $0 $ &    $58 $   ms  \\
 %  \hline
   IMPLICIT GEMM &    $48$ KB  &  $59$   ms \\
  % \hline
   PRECOMP GEMM & $4.8$ GB & $126$ ms \\
  % \hline
   WINOGRAD  NONFUSED & $691$  MB &  $46 $   ms \\
  % \hline
   FFT    & $2.2 $ GB    &   $36 $  ms   \\
  % \hline
   FFT TILING  &   $1.1$ GB & $48$ ms \\
 \hline
\end{tabular}
\vspace{0.4cm}
 \caption{Comparison of workspace memory and execution time for the $5 \times 5$ convolution in the third inception module of GoogleNet on a Tesla K40 GPU using all the algorithms implemented in cuDNN. \texttt{DIRECT} and \texttt{WINOGRAD} algorithms are not supported for this input.}\label{table:ws}
\end{table} 

Thereby, for two convolutions in Table \ref{table:sm}, if we choose PRECOMP\_GEMM for the first convolution and FFT\_TILING for the second (TensorFlow would pick PRECOMP\_GEMM for both) and employ SM partitioning \cite{dai2018accelerate, xu2016warped, zhao2018classification}, the memory stalls of the second convolution can potentially be hidden by switching to compute-warps from the first convolution. 
This parallelization can improve resource utilization and reduce latency compared to serial execution. 
We discover 27 similar cases in this network and more instances in other popular non-linear CNNs such as ResNet. 

\textbf{\emph{Device Memory.}}
Table \ref{table:ws} shows the execution time and workspace memory for a convolution operation in GoogleNet. 
Comparing the FFT algorithm (TensorFlow selection) with Winograd Nonfused, the former is only 21\% faster but requires almost 1.5 GB (or 70\%) of extra memory.
Changing the convolution algorithm is the only way to configure workspace memory. 
%Similar behavior has been observed for resource utilization. 
Therefore, careful and profiling-based algorithm selection has the potential to mitigate concurrent kernel execution's limitations and improve the parallelism on a single GPU.

%For example, backward filter convolution using Winograd Nonfused algorithm in cuDNN underutilizes two-thirds of the SMs for almost half of the execution time.
%Researchers have also reported low DRAM utilization and partition camping in cuDNN forward convolutions using FFT \cite{analyzingML}.
%If cuDNN can accommodate concurrent execution of convolutions, underutilized resources in one convolution can be used in another taking into careful consideration, the algorithm, input data, data layout, and microarchitecture.
\section{Concluding Remarks}
We conclude that partitioning GPU computing resources among concurrent convolutions depends on the workload (algorithm) which impacts both the execution time and workspace memory of kernels. 
While we observe no strong correlation between the execution time and workspace memory, they are mutually dependent.
Moreover, selecting independent operations from the ready queue for concurrent execution is a challenging scheduling problem that highly depends on the network topology and resource utilization of operations. 
Even though we observe the potential for concurrent execution of convolutions, profile-based algorithm selection has to evaluate multiple metrics for optimal parallelism. 

Currently, neither does CUDA provide an API for partitioning compute resources between streams nor does cuDNN API support that configuration.
Therefore, we are investigating open-source frameworks such as AMD ROCm and GPU simulators for implementing intra- and inter-SM partitioning along with profiling-based algorithm selection.

%Unfortunately, neither does CUDA provide an API for partitioning compute resources between streams nor does cuDNN API support that configuration.
%One suggestion for cuDNN developers is to provide a knob in the API for choosing the number of SMs to execute a convolution.
%Partitioning SMs between convolutions alone, however, is not sufficient to launch multiple convolutions concurrently.
%We also have to eliminate resource contention on shared resources (i.e. global memory) between SMs.

%Therefore, we build a model (using a dynamic scheduling method) to pick the optimal algorithm that minimizes train time given resource constraint (e.g. GPU memory). We simulate the model and our initial results show improvement in training time of several contemporary models such as non-linear inception module in GoogleNet.

\bibliographystyle{ACM-Reference-Format}
\balance
\bibliography{main}

%%% -*-BibTeX-*-
%%% Do NOT edit. File created by BibTeX with style
%%% ACM-Reference-Format-Journals [18-Jan-2012].

\begin{thebibliography}{24}

%%% ====================================================================
%%% NOTE TO THE USER: you can override these defaults by providing
%%% customized versions of any of these macros before the \bibliography
%%% command.  Each of them MUST provide its own final punctuation,
%%% except for \shownote{}, \showDOI{}, and \showURL{}.  The latter two
%%% do not use final punctuation, in order to avoid confusing it with
%%% the Web address.
%%%
%%% To suppress output of a particular field, define its macro to expand
%%% to an empty string, or better, \unskip, like this:
%%%
%%% \newcommand{\showDOI}[1]{\unskip}   % LaTeX syntax
%%%
%%% \def \showDOI #1{\unskip}           % plain TeX syntax
%%%
%%% ====================================================================

\ifx \showCODEN    \undefined \def \showCODEN     #1{\unskip}     \fi
\ifx \showDOI      \undefined \def \showDOI       #1{#1}\fi
\ifx \showISBNx    \undefined \def \showISBNx     #1{\unskip}     \fi
\ifx \showISBNxiii \undefined \def \showISBNxiii  #1{\unskip}     \fi
\ifx \showISSN     \undefined \def \showISSN      #1{\unskip}     \fi
\ifx \showLCCN     \undefined \def \showLCCN      #1{\unskip}     \fi
\ifx \shownote     \undefined \def \shownote      #1{#1}          \fi
\ifx \showarticletitle \undefined \def \showarticletitle #1{#1}   \fi
\ifx \showURL      \undefined \def \showURL       {\relax}        \fi
% The following commands are used for tagged output and should be
% invisible to TeX
\providecommand\bibfield[2]{#2}
\providecommand\bibinfo[2]{#2}
\providecommand\natexlab[1]{#1}
\providecommand\showeprint[2][]{arXiv:#2}

\bibitem[\protect\citeauthoryear{??}{cub}{[n. d.]}]%
        {cublas}
 \bibinfo{year}{[n. d.]}\natexlab{}.
\newblock \bibinfo{title}{cuBLAS}.
\newblock
\newblock
\newblock
\shownote{\url{https://developer.nvidia.com/cublas}.}


\bibitem[\protect\citeauthoryear{Abdel-Hamid, Mohamed, Jiang, Deng, Penn, and
  Yu}{Abdel-Hamid et~al\mbox{.}}{2014}]%
        {speech}
\bibfield{author}{\bibinfo{person}{Ossama Abdel-Hamid},
  \bibinfo{person}{Abdel-rahman Mohamed}, \bibinfo{person}{Hui Jiang},
  \bibinfo{person}{Li Deng}, \bibinfo{person}{Gerald Penn}, {and}
  \bibinfo{person}{Dong Yu}.} \bibinfo{year}{2014}\natexlab{}.
\newblock \showarticletitle{Convolutional neural networks for speech
  recognition}.
\newblock \bibinfo{journal}{\emph{IEEE/ACM Transactions on audio, speech, and
  language processing}} \bibinfo{volume}{22}, \bibinfo{number}{10}
  (\bibinfo{year}{2014}), \bibinfo{pages}{1533--1545}.
\newblock


\bibitem[\protect\citeauthoryear{Adriaens, Compton, Kim, and Schulte}{Adriaens
  et~al\mbox{.}}{2012}]%
        {adriaens2012case}
\bibfield{author}{\bibinfo{person}{Jacob~T Adriaens},
  \bibinfo{person}{Katherine Compton}, \bibinfo{person}{Nam~Sung Kim}, {and}
  \bibinfo{person}{Michael~J Schulte}.} \bibinfo{year}{2012}\natexlab{}.
\newblock \showarticletitle{The case for GPGPU spatial multitasking}. In
  \bibinfo{booktitle}{\emph{IEEE International Symposium on High-Performance
  Comp Architecture}}. IEEE, \bibinfo{pages}{1--12}.
\newblock


\bibitem[\protect\citeauthoryear{Ben-Nun and Hoefler}{Ben-Nun and
  Hoefler}{2019}]%
        {ben2019demystifying}
\bibfield{author}{\bibinfo{person}{Tal Ben-Nun} {and} \bibinfo{person}{Torsten
  Hoefler}.} \bibinfo{year}{2019}\natexlab{}.
\newblock \showarticletitle{Demystifying parallel and distributed deep
  learning: An in-depth concurrency analysis}.
\newblock \bibinfo{journal}{\emph{ACM Computing Surveys (CSUR)}}
  \bibinfo{volume}{52}, \bibinfo{number}{4} (\bibinfo{year}{2019}),
  \bibinfo{pages}{1--43}.
\newblock


\bibitem[\protect\citeauthoryear{Chetlur, Woolley, Vandermersch, Cohen, Tran,
  Catanzaro, and Shelhamer}{Chetlur et~al\mbox{.}}{2014}]%
        {cudnn}
\bibfield{author}{\bibinfo{person}{Sharan Chetlur}, \bibinfo{person}{Cliff
  Woolley}, \bibinfo{person}{Philippe Vandermersch}, \bibinfo{person}{Jonathan
  Cohen}, \bibinfo{person}{John Tran}, \bibinfo{person}{Bryan Catanzaro}, {and}
  \bibinfo{person}{Evan Shelhamer}.} \bibinfo{year}{2014}\natexlab{}.
\newblock \showarticletitle{{cuDNN: Efficient primitives for deep learning}}.
\newblock \bibinfo{journal}{\emph{arXiv preprint arXiv:1410.0759}}
  (\bibinfo{year}{2014}).
\newblock


\bibitem[\protect\citeauthoryear{Collobert and Weston}{Collobert and
  Weston}{2008}]%
        {nlp2}
\bibfield{author}{\bibinfo{person}{Ronan Collobert} {and}
  \bibinfo{person}{Jason Weston}.} \bibinfo{year}{2008}\natexlab{}.
\newblock \showarticletitle{A unified architecture for natural language
  processing: Deep neural networks with multitask learning}. In
  \bibinfo{booktitle}{\emph{Proceedings of the 25th international conference on
  Machine learning}}. \bibinfo{pages}{160--167}.
\newblock


\bibitem[\protect\citeauthoryear{Dai, Lin, Li, Zhao, Wang, Zheng, and Zhou}{Dai
  et~al\mbox{.}}{2018}]%
        {dai2018accelerate}
\bibfield{author}{\bibinfo{person}{Hongwen Dai}, \bibinfo{person}{Zhen Lin},
  \bibinfo{person}{Chao Li}, \bibinfo{person}{Chen Zhao}, \bibinfo{person}{Fei
  Wang}, \bibinfo{person}{Nanning Zheng}, {and} \bibinfo{person}{Huiyang
  Zhou}.} \bibinfo{year}{2018}\natexlab{}.
\newblock \showarticletitle{Accelerate GPU concurrent kernel execution by
  mitigating memory pipeline stalls}. In \bibinfo{booktitle}{\emph{2018 IEEE
  International Symposium on High Performance Computer Architecture (HPCA)}}.
  IEEE, \bibinfo{pages}{208--220}.
\newblock


\bibitem[\protect\citeauthoryear{Jia, Zaharia, and Aiken}{Jia
  et~al\mbox{.}}{2018}]%
        {jia2018beyond}
\bibfield{author}{\bibinfo{person}{Zhihao Jia}, \bibinfo{person}{Matei
  Zaharia}, {and} \bibinfo{person}{Alex Aiken}.}
  \bibinfo{year}{2018}\natexlab{}.
\newblock \showarticletitle{Beyond data and model parallelism for deep neural
  networks}.
\newblock \bibinfo{journal}{\emph{arXiv preprint arXiv:1807.05358}}
  (\bibinfo{year}{2018}).
\newblock


\bibitem[\protect\citeauthoryear{Kim}{Kim}{2014}]%
        {nlp}
\bibfield{author}{\bibinfo{person}{Yoon Kim}.} \bibinfo{year}{2014}\natexlab{}.
\newblock \showarticletitle{Convolutional neural networks for sentence
  classification}.
\newblock \bibinfo{journal}{\emph{arXiv preprint arXiv:1408.5882}}
  (\bibinfo{year}{2014}).
\newblock


\bibitem[\protect\citeauthoryear{Lavin}{Lavin}{2015}]%
        {maxdnn}
\bibfield{author}{\bibinfo{person}{Andrew Lavin}.}
  \bibinfo{year}{2015}\natexlab{}.
\newblock \showarticletitle{{maxDNN}: {An} efficient convolution kernel for
  deep learning with maxwell {GPUs}}.
\newblock \bibinfo{journal}{\emph{arXiv preprint arXiv:1501.06633}}
  (\bibinfo{year}{2015}).
\newblock


\bibitem[\protect\citeauthoryear{Lavin and Gray}{Lavin and Gray}{2016}]%
        {lavin2016fast}
\bibfield{author}{\bibinfo{person}{Andrew Lavin} {and} \bibinfo{person}{Scott
  Gray}.} \bibinfo{year}{2016}\natexlab{}.
\newblock \showarticletitle{Fast algorithms for convolutional neural networks}.
  In \bibinfo{booktitle}{\emph{Proc. of the IEEE Conf. on Computer Vision and
  Pattern Recognition}}. \bibinfo{pages}{4013--4021}.
\newblock


\bibitem[\protect\citeauthoryear{LeCun, Kavukcuoglu, Farabet,
  et~al\mbox{.}}{LeCun et~al\mbox{.}}{2010}]%
        {vision}
\bibfield{author}{\bibinfo{person}{Yann LeCun}, \bibinfo{person}{Koray
  Kavukcuoglu}, \bibinfo{person}{Cl{\'e}ment Farabet}, {et~al\mbox{.}}}
  \bibinfo{year}{2010}\natexlab{}.
\newblock \showarticletitle{Convolutional networks and applications in
  vision.}. In \bibinfo{booktitle}{\emph{ISCAS}}, Vol.~\bibinfo{volume}{2010}.
  \bibinfo{pages}{253--256}.
\newblock


\bibitem[\protect\citeauthoryear{Lew, Shah, Pati, Cattell, Zhang, Sandhupatla,
  Ng, Goli, Sinclair, Rogers, and Aamodt}{Lew et~al\mbox{.}}{2018}]%
        {analyzingML}
\bibfield{author}{\bibinfo{person}{Jonathan Lew}, \bibinfo{person}{Deval Shah},
  \bibinfo{person}{Suchita Pati}, \bibinfo{person}{Shaylin Cattell},
  \bibinfo{person}{Mengchi Zhang}, \bibinfo{person}{Amruth Sandhupatla},
  \bibinfo{person}{Christopher Ng}, \bibinfo{person}{Negar Goli},
  \bibinfo{person}{Matthew~D Sinclair}, \bibinfo{person}{Timothy~G Rogers},
  {and} \bibinfo{person}{Tor Aamodt}.} \bibinfo{year}{2018}\natexlab{}.
\newblock \showarticletitle{Analyzing Machine Learning Workloads Using a
  Detailed {GPU} Simulator}.
\newblock \bibinfo{journal}{\emph{arXiv preprint arXiv:1811.08933}}
  (\bibinfo{year}{2018}).
\newblock


\bibitem[\protect\citeauthoryear{Li, Yang, Feng, Chakradhar, and Zhou}{Li
  et~al\mbox{.}}{2016}]%
        {li2016optimizing}
\bibfield{author}{\bibinfo{person}{Chao Li}, \bibinfo{person}{Yi Yang},
  \bibinfo{person}{Min Feng}, \bibinfo{person}{Srimat Chakradhar}, {and}
  \bibinfo{person}{Huiyang Zhou}.} \bibinfo{year}{2016}\natexlab{}.
\newblock \showarticletitle{Optimizing memory efficiency for deep convolutional
  neural networks on {GPUs}}. In \bibinfo{booktitle}{\emph{Proc. of the Int'l
  Conf. for High Performance Computing, Networking, Storage and Analysis}}.
  IEEE, \bibinfo{pages}{633--644}.
\newblock


\bibitem[\protect\citeauthoryear{Obiols-Sales, Vishnu, Malaya, and
  Chandramowlishwaran}{Obiols-Sales et~al\mbox{.}}{2020}]%
        {obiols2020cfdnet}
\bibfield{author}{\bibinfo{person}{Octavi Obiols-Sales},
  \bibinfo{person}{Abhinav Vishnu}, \bibinfo{person}{Nicholas Malaya}, {and}
  \bibinfo{person}{Aparna Chandramowlishwaran}.}
  \bibinfo{year}{2020}\natexlab{}.
\newblock \showarticletitle{CFDNet: A deep learning-based accelerator for fluid
  simulations}.
\newblock \bibinfo{journal}{\emph{arXiv preprint arXiv:2005.04485}}
  (\bibinfo{year}{2020}).
\newblock


\bibitem[\protect\citeauthoryear{Park, Park, and Mahlke}{Park
  et~al\mbox{.}}{2017}]%
        {park2017dynamic}
\bibfield{author}{\bibinfo{person}{Jason Jong~Kyu Park},
  \bibinfo{person}{Yongjun Park}, {and} \bibinfo{person}{Scott Mahlke}.}
  \bibinfo{year}{2017}\natexlab{}.
\newblock \showarticletitle{Dynamic resource management for efficient
  utilization of multitasking GPUs}. In \bibinfo{booktitle}{\emph{Proceedings
  of the Twenty-Second International Conference on Architectural Support for
  Programming Languages and Operating Systems}}. \bibinfo{pages}{527--540}.
\newblock


\bibitem[\protect\citeauthoryear{Simonyan and Zisserman}{Simonyan and
  Zisserman}{2014}]%
        {vgg}
\bibfield{author}{\bibinfo{person}{Karen Simonyan} {and}
  \bibinfo{person}{Andrew Zisserman}.} \bibinfo{year}{2014}\natexlab{}.
\newblock \showarticletitle{Very deep convolutional networks for large-scale
  image recognition}.
\newblock \bibinfo{journal}{\emph{arXiv preprint arXiv:1409.1556}}
  (\bibinfo{year}{2014}).
\newblock


\bibitem[\protect\citeauthoryear{Strigl, Kofler, and Podlipnig}{Strigl
  et~al\mbox{.}}{2010}]%
        {cnnperformance}
\bibfield{author}{\bibinfo{person}{Daniel Strigl}, \bibinfo{person}{Klaus
  Kofler}, {and} \bibinfo{person}{Stefan Podlipnig}.}
  \bibinfo{year}{2010}\natexlab{}.
\newblock \showarticletitle{Performance and scalability of {GPU-based}
  convolutional neural networks}. In \bibinfo{booktitle}{\emph{18th Euromicro
  International Conf. on Parallel, Distributed and Network-Based Processing}}.
  IEEE, \bibinfo{pages}{317--324}.
\newblock


\bibitem[\protect\citeauthoryear{Tang, Wang, Willke, and Li}{Tang
  et~al\mbox{.}}{2018}]%
        {graphi}
\bibfield{author}{\bibinfo{person}{Linpeng Tang}, \bibinfo{person}{Yida Wang},
  \bibinfo{person}{Theodore~L Willke}, {and} \bibinfo{person}{Kai Li}.}
  \bibinfo{year}{2018}\natexlab{}.
\newblock \showarticletitle{Scheduling Computation Graphs of Deep Learning
  Models on Manycore {CPUs}}.
\newblock \bibinfo{journal}{\emph{arXiv preprint arXiv:1807.09667}}
  (\bibinfo{year}{2018}).
\newblock


\bibitem[\protect\citeauthoryear{Wang, Ye, Zhao, Wu, Li, Song, Xu, and
  Kraska}{Wang et~al\mbox{.}}{2018}]%
        {superneurons}
\bibfield{author}{\bibinfo{person}{Linnan Wang}, \bibinfo{person}{Jinmian Ye},
  \bibinfo{person}{Yiyang Zhao}, \bibinfo{person}{Wei Wu}, \bibinfo{person}{Ang
  Li}, \bibinfo{person}{Shuaiwen~Leon Song}, \bibinfo{person}{Zenglin Xu},
  {and} \bibinfo{person}{Tim Kraska}.} \bibinfo{year}{2018}\natexlab{}.
\newblock \showarticletitle{{Superneurons: Dynamic GPU} memory management for
  training deep neural networks}. In \bibinfo{booktitle}{\emph{Proc. of the ACM
  SIGPLAN Symp. on Principles and Practice of Parallel Programming}}. ACM,
  \bibinfo{pages}{41--53}.
\newblock


\bibitem[\protect\citeauthoryear{Wang, Yang, Melhem, Childers, Zhang, and
  Guo}{Wang et~al\mbox{.}}{2016}]%
        {wang2016simultaneous}
\bibfield{author}{\bibinfo{person}{Zhenning Wang}, \bibinfo{person}{Jun Yang},
  \bibinfo{person}{Rami Melhem}, \bibinfo{person}{Bruce Childers},
  \bibinfo{person}{Youtao Zhang}, {and} \bibinfo{person}{Minyi Guo}.}
  \bibinfo{year}{2016}\natexlab{}.
\newblock \showarticletitle{Simultaneous multikernel GPU: Multi-tasking
  throughput processors via fine-grained sharing}. In
  \bibinfo{booktitle}{\emph{2016 IEEE International Symposium on High
  Performance Computer Architecture (HPCA)}}. IEEE, \bibinfo{pages}{358--369}.
\newblock


\bibitem[\protect\citeauthoryear{Xu, Jeon, Kim, Ro, and Annavaram}{Xu
  et~al\mbox{.}}{2016}]%
        {xu2016warped}
\bibfield{author}{\bibinfo{person}{Qiumin Xu}, \bibinfo{person}{Hyeran Jeon},
  \bibinfo{person}{Keunsoo Kim}, \bibinfo{person}{Won~Woo Ro}, {and}
  \bibinfo{person}{Murali Annavaram}.} \bibinfo{year}{2016}\natexlab{}.
\newblock \showarticletitle{Warped-slicer: efficient intra-SM slicing through
  dynamic resource partitioning for GPU multiprogramming}. In
  \bibinfo{booktitle}{\emph{2016 ACM/IEEE 43rd Annual International Symposium
  on Computer Architecture (ISCA)}}. IEEE, \bibinfo{pages}{230--242}.
\newblock


\bibitem[\protect\citeauthoryear{Ying, He, Chen, Eksombatchai, Hamilton, and
  Leskovec}{Ying et~al\mbox{.}}{2018}]%
        {recommender}
\bibfield{author}{\bibinfo{person}{Rex Ying}, \bibinfo{person}{Ruining He},
  \bibinfo{person}{Kaifeng Chen}, \bibinfo{person}{Pong Eksombatchai},
  \bibinfo{person}{William~L Hamilton}, {and} \bibinfo{person}{Jure Leskovec}.}
  \bibinfo{year}{2018}\natexlab{}.
\newblock \showarticletitle{Graph convolutional neural networks for web-scale
  recommender systems}. In \bibinfo{booktitle}{\emph{Proceedings of the 24th
  ACM SIGKDD International Conference on Knowledge Discovery \& Data Mining}}.
  \bibinfo{pages}{974--983}.
\newblock


\bibitem[\protect\citeauthoryear{Zhao, Wang, and Eeckhout}{Zhao
  et~al\mbox{.}}{2018}]%
        {zhao2018classification}
\bibfield{author}{\bibinfo{person}{Xia Zhao}, \bibinfo{person}{Zhiying Wang},
  {and} \bibinfo{person}{Lieven Eeckhout}.} \bibinfo{year}{2018}\natexlab{}.
\newblock \showarticletitle{Classification-driven search for effective sm
  partitioning in multitasking GPUs}. In \bibinfo{booktitle}{\emph{Proceedings
  of the 2018 International Conference on Supercomputing}}.
  \bibinfo{pages}{65--75}.
\newblock


\end{thebibliography}

\end{document}